\documentclass[twocolumn]{revtex4}

\usepackage{tikz}
\usetikzlibrary{arrows.meta,decorations.markings,calc}
\usepackage{amssymb} 

\usepackage[english]{babel} 
\usepackage{bm}
\usepackage{graphicx}
\usepackage{amsmath}
\usepackage{amssymb}
\usepackage{color}
\usepackage{comment}
\usepackage[hidelinks]{hyperref}
\usepackage{version}

\DeclareMathOperator{\tr}{tr}
\DeclareMathOperator{\sgn}{sgn}
	
\begin{document}

\excludeversion{details}
		
		\title{Electric excitation of spin resonance 
        in altermagnetic and antiferromagnetic conductors} 
		
		\author{R. Ramazashvili}
		\email[]{revaz.ramazashvili@cnrs.fr}
		\affiliation{Univ Toulouse, CNRS, LPT, Toulouse, France}
        \author{V. Shablenko}
		\email[]{shablenv@email.sc.edu}
		\affiliation{Department of Physics \& Astronomy, University of South Carolina, Columbia, South Carolina 29208, USA}
        \author{Ya. B. Bazaliy}
		\email[]{bazaliy@mailbox.sc.edu}
		\affiliation{Department of Physics \& Astronomy, University of South Carolina, Columbia, South Carolina 29208, USA}
		
		\date{\today}
		
		\begin{abstract} 
        We predict electric-dipole spin resonance (EDSR) in altermagnetic conductors: 
        in a magnetic field ${\mathbf H}_\perp$, perpendicular to the magnetization axis, 
        an AC electric field will induce a \textit{spin} resonance peak above the lower 
        threshold frequency $\omega_- = 2H_\perp$. 
        In sufficiently clean samples, 
        this peak shall be clearly visible on the background of ohmic absorption. 
        EDSR can thus serve as a diagnostic of altermagnetism in conducting materials 
        and as a means to distinguish it from higher-symmetry antiferromagnetic order.
		\end{abstract}
		
		\maketitle
		
Magnetic order in solids finds itself in an enviable position: on the one hand, it is often present in strongly correlated materials such as cuprate and organic superconductors, as well as Moir\'e graphene systems, that continue to attract considerable attention because of the fundamental questions they raise. At the same time, magnetic compounds hold promise as materials of choice for the next generation of devices that will employ electron spin and charge on an equal footing. In this context, a new class of antiferromagnetic materials, dubbed altermagnets, 
has recently sparked enormous interest owing to momentum-dependent spin 
splitting of electron bands, occurring 
even in the complete absence of relativistic spin–orbit coupling \cite{Hayami_2019,Smejkal_2020,Yuan_2020}. 
Various materials are being scrutinized as altermagnet candidates -- most often, 
with the help of spin- and angle-resolved photoemission spectroscopy \cite{Ding_2024,Reimers_2024,Zeng_2024,Krempasky_2024,Lee_2024,Osumi_2024}. High energy and 
momentum resolution of such experiments require large synchrotron facilities, 
%
%
%
%
%
%
%
%
%
and it is thus desirable to find simpler tools that could allow one to diagnose altermagnetism
-- and distinguish it from antiferromagnets with degenerate electron bands.

Here we show that momentum-dependent spin splitting produces a distinctive experimental signature and diagnostic for antiferromagnetic conductors in general, and for altermagnetic conductors in particular. It enables electric-dipole spin resonance (EDSR) \cite{Rashba-Sheka-1991}: \textit{electrically} driven spin-flip transitions. 
In a static magnetic field transverse to the magnetization axis, 
EDSR generates a peaked spin contribution to the absorption rate that, 
in sufficiently clean samples or/and at sufficiently high fields, 
should stand out clearly from the slowly varying background 
of ohmic absorption. 

Motivated by $d$-wave band splitting, pointed out for electron-doped 
quasi-two-dimensional antiferromagnetic organic salts 
$\kappa$-(BEDT-TTF)$_2$X \cite{Naka_2019}, 
by recently identified mixed-valence layered altermagnetic 
candidate materials such as vanadium oxyselenide KV$_2$Se$_2$O 
\cite{Jiang_2025} and vanadium oxychalcogenide 
Rb$_{1 - \delta}$V$_2$Te$_2$O \cite{Zhang_2025}, 
as well as by ongoing discussion of altermagnetism with 
higher-harmonic splitting in layered materials \cite{Sprague_2026}, 
we illustrate the effect by a minimal model of a two-dimensional 
low-carrier-density $d$-wave altermagnetic conductor. 
Our results provide an independent approach that may help resolve recent controversies between ARPES results \cite{Jiang_2025,Zhang_2025} interpreted as signatures of bulk altermagnetism and neutron-diffraction data \cite{Sun_2025,Xie_2026} revealing bulk magnetic structures that challenge these interpretations in some candidate materials.

The low-energy `altermagnetic' electron Hamiltonian 
reads \cite{Hayami_2019,Smejkal_2022} 
\begin{equation}
\label{eq:H}
    \mathcal{H} = \frac{{\bf p}^2}{2m}
    + \Delta_{\bf p} \sigma_z - \left( {\mathbf H}_\perp \cdot {\bm \sigma} \right).
\end{equation}
Here \mbox{$\Delta_{\bf p} = \frac{p_x^2 - p_y^2}{2M}$} with line nodes 
$p_x = \pm p_y$ describes spontaneous spin splitting, an external magnetic 
field ${\bf H}_\perp$ is applied transversely to the N\'eel  axis $\hat{z}$ 
\footnote{To simplify notation, we absorbed the Bohr magneton $\mu_B$ and 
the $g$-factor into the definition of the field as per 
${\bf H}_\perp = \frac{\mu_B}{2} g {\bf B}_\perp$.},  
and ${\bm \sigma} = (\sigma_x, \sigma_y, \sigma_z)$ is the triad of spin Pauli matrices.  
We make no assumption regarding the orientation of the 
N\'eel axis $\hat{z}$ relative to the conducting plane. 
The two bands 
$\mathcal{E}_\pm = \frac{{\bf p}^2}{2m} \pm \sqrt{\Delta_{\bf p}^2 + H_\perp^2}$  
of Hamiltonian (\ref{eq:H}) correspond to `up' and `down' spin projections 
on the axis, defined by unit vector 
${\bm s}_{\bf p}^{(0)} = (\Delta_{\bf p} \hat{z} - {\bf H}_\perp)/ 
\sqrt{\Delta_{\bf p}^2 + H_\perp^2}$ -- 
and have momentum-dependent splitting  
$\hbar \omega_{\bf p} = 2 \sqrt{\Delta_{\bf p}^2 + H_\perp^2}$. 
The ratio $m/M$ quantifies the strength of the splitting term 
$\Delta_{\bf p} \sigma_z$ relative to that of the main dispersion term 
$\frac{{\bf p}^2}{2m}$.
Both spin-split subbands having a minimum 
at \mbox{${\bf p} = 0$} requires $m/M < 1$; hereafter we assume $m/M \ll 1$. 
This captures the physics while simplifying the analysis -- and provides 
a good first approximation for the materials studied in Refs. \cite{Naka_2019,Jiang_2025,Zhang_2025}. 
Finally, we completely neglect the coupling of the magnetic field 
to the orbital motion of the electron. In a layered compound, this implies 
${\bf H}_\perp$ pointing along the layers.

In an experiment, a uniform AC electric field ${\bf E}$ is applied along the layers. 
The ${\bf E}$ is expressed via the vector potential ${\bf A}$ 
as ${\bf E} = - \frac{1}{c} \partial_t {\bf A}$, with the ${\bf A}$ 
entering Hamiltonian (\ref{eq:H}) through the Peierls substitution 
 ${\bf p} \rightarrow {\bf p} - \frac{e}{c}{\bf A}$, 
 where $e$ is the electron charge, and $c$ the speed of light. 
To first order in $\delta {\bf A}$, the perturbation is
\begin{equation}
\label{eq:dH}
    \delta \mathcal{H}_{\bf p} = - \frac{1}{c} (\delta{\bf A} \cdot {\bf j})  
    = - \frac{e}{c} \delta{\bf A} \nabla_{\bf p} \mathcal{H}  
    = \delta \mathcal{H}_{\bf p}^e + \delta \mathcal{H}_{\bf p}^s ,
\end{equation}
with the familiar charge-current term  
$\delta \mathcal{H}_{\bf p}^e =
-\nobreak\frac{1}{c}\nobreak(\delta\nobreak{\bf A}\nobreak\cdot\nobreak{\bf j}^e) 
= - \frac{1}{c} \delta{\bf A} \cdot \frac{e}{m} 
\left[ {\bf p} - \frac{e}{c} {\bf A} \right]$ 
and its spin counterpart 
$\delta \mathcal{H}_{\bf p}^s = - \frac{1}{c} (\delta{\bf A} \cdot {\bf j}^s)
= - \frac{1}{c} \delta{\bf A} \cdot e \sigma_z 
\frac{\partial}{\partial {\bf p}} \Delta_{{\bf p} - \frac{e}{c} {\bf A}}$. 
As a result, the AC field induces two kinds of transitions, 
and hence two distinct contributions to the absorption. 
The $\delta \mathcal{H}_{\bf p}^e$ in Eq. (\ref{eq:dH}) is inert with respect to 
spin and gives rise to textbook electric conduction, 
while $\delta \mathcal{H}_{\bf p}^s$ produces EDSR: ``vertical'' transitions between 
the spin-split eigenstates of Hamiltonian (\ref{eq:H}) -- the focus of this work. 

Direct comparison shows that the matrix elements 
of the $\delta \mathcal{H}_{\bf p}^s$ are suppressed 
with respect to those of $\delta \mathcal{H}_{\bf p}^e$ 
by about $m/M \ll 1$ -- and thus, at first sight, the EDSR 
absorption shall make only a tiny contribution of the order 
of $(m/M)^2 \lll 1$ relative to the ohmic dissipation. 
However, this reasoning is misleading, 
as can be seen already from the Fermi’s golden rule \cite{LL-III} expression 
for the EDSR absorption rate $P_\omega^s$ in a perfect crystal:
\begin{equation}
\label{eq:Fermi}
    P_\omega^s = 
    \frac{2 \pi \omega}{\hbar} \sum_{\bf p} | \delta \mathcal{H}_{\bf p}^s |^2 
 \delta \left( \omega - \omega_{\bf p} \right) ,
\end{equation}
where $| \delta \mathcal{H}_{\bf p} |^2$ is the squared modulus of the matrix 
element of $\delta \mathcal{H}_{\bf p}^s$ between the eigenstates of Hamiltonian (\ref{eq:H}). 
Note that, in Eq. (\ref{eq:Fermi}), 
the band splitting $\hbar\omega_p$, defined below Eq.~(\ref{eq:H}), 
has a flat minimum $\hbar \omega_{\bf p} = 2 H_\perp$ 
along the nodal lines of $\Delta_{\bf p}$. As a result, the $\mathcal{P}_\omega^s$ 
has a lower threshold at \mbox{$\omega_- = 2 H_\perp / \hbar$}. 
Above this threshold, the density of states 
$\nu (\omega) = \sum_{\bf p} \delta \left( \omega - \omega_{\bf p} \right)$ 
of $\hbar \omega_{\bf p}$ diverges as \mbox{$(\omega - \omega_-)^{-1/2}$}, 
which overwhelms the apparent suppression of the EDSR matrix elements. For the EDSR absorption rate $\mathcal{P}_\omega^s$ per unit volume, we obtain 
\begin{equation}
    \label{eq:main-peak-collisionless}
 \mathcal{P}_\omega^{s}
   = 
   \frac{m}{M} 
   \frac{( e E_\omega )^2}{4 \pi \hbar}
   \frac{\left( \frac{\omega_-}{\omega} \right) }{\sqrt{\left( \frac{\omega}{\omega_-} \right)^2 - 1}} 
    .
\end{equation}

\begin{figure}[b]
    \centering
    \includegraphics[width = 1 \linewidth]{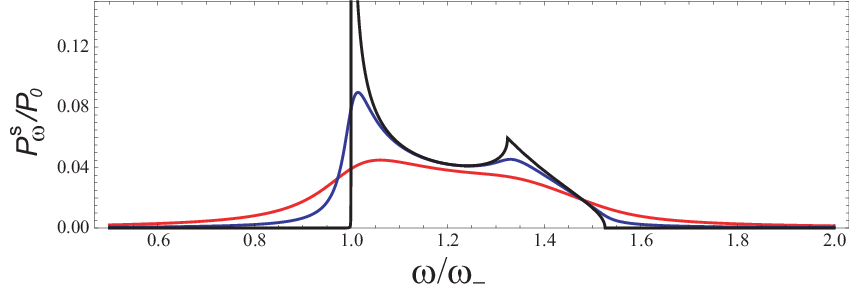}
    \caption{Spin absorption $\mathcal{P}_\omega^{s}$, 
    plotted in units of $\mathcal{P}_0$ 
    (defined immediately below Eq. 
    (\ref{eq:Drude})) for \mbox{$m/M = 0.1$}. 
    The black line represents the clean limit ($\tau = \infty$) 
    and shows the divergent threshold peak of Eq. (\ref{eq:main-peak-collisionless}) 
    and the second, finite peak of Eq. (\ref{eq:geopeak}). 
    In the presence of a finite relaxation time $\tau$, 
    both peaks are rounded off, as illustrated by the blue line ($\omega_{-} \tau = 40$) 
    and the red line ($\omega_{-} \tau = 10$).}
    \label{fig:Ps_family}
\end{figure}

The key features of the EDSR absorption are shown in Fig.~\ref{fig:Ps_family} (black curve). They can be read off the $\nu (\omega)$: 
both $\nu (\omega)$ and $\mathcal{P}_\omega^s$ 
have an upper threshold $\hbar \omega_+$
and exhibit a second---finite rather than divergent---peak at $\hbar\omega_g$. 
For sufficiently weak fields ($\omega_- \ll \mu m / M$) we find 
$\hbar \omega_+ = 2 \mu \frac{m/M}{1 - m/M}$
and $\hbar \omega_g = 2 \mu \frac{m/M}{1 + \frac{m}{M}}$, 
where $\mu$ is the chemical potential of the dopant carriers relative to the bottom of the band. 
Immediately below $\omega_g$ ($\omega_g - \omega \ll \omega_g$), we find 
\begin{equation}
    \label{eq:geopeak}
    \mathcal{P}_{\omega}^s 
    \approx
\frac{( e E_\omega)^2}{4 \pi \hbar} 
\sqrt{\frac{m}{M}}
\left( \frac{\omega_-}{\omega_g} \right)^2  
\left[
1 - \sqrt{\frac{M}{m}} \sqrt{\frac{\omega_g - \omega}{2 \omega_g}}
\right]  .
\end{equation}

A more complete yet transparent description of the EDSR 
is afforded by the kinetic equation, to which we now turn. 
To find the absorption rate 
$\mathcal{P}_\omega = \int d^2 {\bf r} ({\bf E}_\omega \cdot {\bf j})$, 
we need to calculate the current ${\bf j}$ as per 
\begin{equation}
\label{eq:j}
    {\bf j} = e \tr \int \frac{d^2 {\bf p}}{(2 \pi \hbar)^2} \hat{\bf v} \hat\rho , 
\end{equation}
with the velocity operator $\hat{\bf v} = {\bm \nabla}_{\bf p} \mathcal{H}$ 
determined by Hamiltonian (\ref{eq:H}):
\begin{equation}
\label{eq:v}
    \hat{\bf v} 
  = \frac{\bf p}{m} +  \sigma_z \nabla_{\bf p} \Delta_{\bf p} ,
\end{equation}
and $\hat\rho$ the density matrix. 
To define the latter, recall that, in equilibrium, 
the filling of the two bands $\mathcal{E}_\pm$ of 
Hamiltonian (\ref{eq:H}) is described by the Fermi distribution 
functions $n^\pm_{eq} = n_F (\mathcal{E}_\pm)$. 
These are the eigenvalues of 
\begin{equation}
\label{eq:rho}
    \hat\rho_{eq} = \frac{1}{2} \left[ n^+_{eq} + n^-_{eq} \right] + 
    \frac{1}{2} \left[ n^+_{eq} - n^-_{eq} \right] ({\bm s}_{\bf p}^{(0)} \cdot {\bm \sigma}) 
\end{equation} 
with ${\bm s}_{\bf p}^{(0)}$ defined below Eq. (\ref{eq:H}) -- and, 
of course, with $\hat\rho = \hat\rho_{eq}$ 
Eq. (\ref{eq:j}) yields zero current and zero absorption. 
A weak AC field ${\bf E}$ perturbs the density matrix away 
from its equilibrium configuration $\hat\rho_{eq}$, and turns 
the $n^\pm_{eq}$ and ${\bm s}_{\bf p}^{(0)}$ into dynamic variables 
$n^\pm$ and ${\bm s}_{\bf p}$ that parameterize the $\hat\rho$. 
As a result, in the linear-response regime, the induced current will involve 
two distinct contributions. One is linear in $\delta n^\pm = n^\pm - n^\pm_{eq}$ 
and yields the familiar ohmic conduction \cite{Abrikosov-fundamentals}, 
while the other is linear in the perturbation $\delta {\bm s}_{\bf p}$ 
of the ${\bm s}_{\bf p}^{(0)}$ -- and is responsible for the electric 
excitation of spin resonance that we will now consider. 

In a uniform AC electric field ${\bf E}_\omega$, the spatial gradients of the 
density matrix vanish, and the kinetic equation reads \cite{Abrikosov-fundamentals,White2007}
\begin{equation}
    \label{eq:Boltzmann-general}
 \left( \partial_t  + e {\bf E}_\omega \cdot \nabla_{\bf p} \right) \hat\rho
+ \frac{i}{\hbar} \left[ \mathcal{H} ,  \hat\rho \right]  = \hat{\mathcal{I}}[\hat\rho] ,
\end{equation} 
with the collision integral in the right-hand side. 
Limiting ourselves to the relaxation-time approximation, 
we find the equation for the ${\bm s}_{\bf p}$:
\begin{equation} 
\label{eq:Boltzmann-transverse}
 \left(  \partial_t  + e {\bf E}_\omega \cdot \nabla_{\bf p} \right) {\bm s}_{\bf p} 
 - \frac{2}{\hbar} \left( {\bm \Delta}_{\bf p} - {\bf H}_\perp \right) \times {\bm s}_{\bf p} 
  = - \frac{1}{\tau_s} \left[ {\bm s}_{\bf p} - {\bm s}_{\bf p}^{(0)} \right] ,
\end{equation} 
where $\tau_s$ is the relaxation time. 
The static solution of Eq. (\ref{eq:Boltzmann-transverse}) 
in the absence of the driving field ${\bf E}_\omega$ is 
${\bm s}_{\bf p}^{(0)} \| \left( {\bm \Delta}_{\bf p} - {\bf H}_\perp \right)$. 
Moreover, since ${\bm s}_{\bf p}$ is a unit vector, one finds 
$({\bm s}_{\bf p} \cdot \partial_t {\bm s}_{\bf p}) = 0$ and 
$({\bm s}_{\bf p} \cdot \nabla_{\bf p} {\bm s}_{\bf p}) = 0$.  
All terms in the l.h.s. are thus normal to the local 
${\bm s}_{\bf p}^{(0)}$ -- and, therefore, 
Eq. (\ref{eq:Boltzmann-transverse}) describes the 
AC field-driven precession of the ${\bm s}_{\bf p}$ 
about its local equilibrium value. 

Eq. (\ref{eq:Boltzmann-transverse}) can be solved in the linear-response 
approximation \cite{Abrikosov-fundamentals,White2007}, with the dominant 
contribution to absorption per unit volume given by 
\begin{equation}
\label{eq:absorption.w.scattering-general-YaB}
   \mathcal{P}_\omega^{s}
 = \frac{1}{2\hbar}
 \int \frac{d^2 {\bf p}}{(2 \pi \hbar)^2} 
 \left[ n^{-}_{eq} - n^{+}_{eq} \right]
 \frac{\omega_-^2}{\omega_{\bf p}^3 \tau_s} 
 \frac{ ( e {\bf E}_\omega \cdot \nabla_{\bf p} \Delta_{\bf p} )^2  }{( \omega   -  \omega_{\bf p})^2 + \tau_s^{-2}} ,
\end{equation} 
where the $\left[ n^{-}_{eq} - n^{+}_{eq} \right]$ accounts for the resonant transitions being possible only from the filled states in the lower subband $\mathcal{E}_-$ to the empty states in the upper subband $\mathcal{E}_+$. By way of a sanity check, we recast 
the $\delta \mathcal{H}_{\bf p}^s$ of Eq. (\ref{eq:dH}) as 
\mbox{$\delta \mathcal{H}_{\bf p}^s = \frac{i}{\omega} \left(e {\bf E}_\omega \cdot \nabla_{\bf p} \Delta_{\bf p} \right)  \sigma_z$} -- and observe that, in the limit of 
vanishing relaxation ($\tau_s \rightarrow \infty$), Eq. (\ref{eq:absorption.w.scattering-general-YaB}) turns into Eq. (\ref{eq:Fermi}) -- and describes the threshold peak  
of Eq. (\ref{eq:main-peak-collisionless}), as it should. 

A finite $\tau_s$ in Eq. (\ref{eq:absorption.w.scattering-general-YaB}) 
cuts off the threshold divergence of Eq. (\ref{eq:main-peak-collisionless}), 
rounding it into a smooth finite peak -- and also turns 
the $\sqrt{\omega_g - \omega}$ cusp at $\omega = \omega_g$ 
in Eq. (\ref{eq:geopeak}) into a smooth maximum, as shown in Fig. \ref{fig:Ps_family}. 
However, recall that these features appear 
on the background of the ohmic absorption 
\begin{equation}
\label{eq:Drude}
\mathcal{P}_{\omega}^e 
 = 
    \sigma_D (\omega) E_\omega^2 = \frac{\sigma_0 E_\omega^2}{1 + (\omega \tau)^2}  
 = 
 \mathcal{P}_0 \frac{\mu \tau / \hbar}{1 + (\omega \tau)^2} ,
\end{equation}
with $\mathcal{P}_0 = \frac{( e E_\omega )^2}{\pi \hbar} = G_0 E_\omega^2$ 
being the absorption, corresponding to the conductance quantum $G_0 = \frac{e^2}{\pi \hbar}$,
as follows from the kinetic equation for the scalar parts 
$n_+$ and $n_-$ of the density matrix \cite{Abrikosov-fundamentals}. 
In Eq. (\ref{eq:Drude}), $\tau$ is the transport scattering time, 
$\sigma_0 = \frac{n e^2 \tau}{m}$ is the DC Drude conductivity and 
$n$ is the electron concentration. 

As a result, the two EDSR features are discernible on the background 
of the $\mathcal{P}_{\omega}^e$ only for $k_F l \approx 2 \mu \tau \gtrsim 300$, 
a bound that may be unattainable for the current generation of samples. 
Yet the sharp increase of $\mathcal{P}_\omega^{s}$ as $\omega$ 
approaches $\omega_-$ from below is robust and clearly visible 
even for much smaller values of $k_F l$. 
The analysis of the r.h.s. of Eq. (\ref{eq:absorption.w.scattering-general-YaB}) 
yields an approximate expression 
\begin{equation}
\label{eq:threshold-YaB}
    \mathcal{P}_\omega^{s}
 = 
  \frac{m}{M} 
  \frac{\mathcal{P}_0}{8} 
 \sqrt{\omega_- \tau_s} J(u) ,
\end{equation}
where 
$u = \tau_s (\omega - \omega_-)$, and $J(x) = \sqrt{\frac{x + \sqrt{x^2 + 1}}{x^2 + 1}}$ 
has a sharp maximum $J(x_m) \approx 1.14$ at $x_m = 1/ \sqrt{3}$.

\begin{figure}[t]
    \centering
    \includegraphics[width = 1.0 \linewidth]{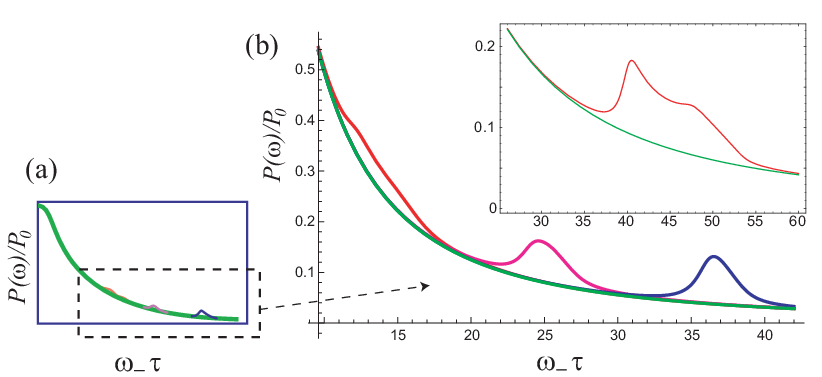}
    \caption{ (a) Sketch of the full absorption
    \mbox{$\mathcal{P} (\omega) = \mathcal{P}_\omega^e + \mathcal{P}_{\omega}^s$} (not to scale), showing the Drude background (green line) and the EDSR peaks against its backdrop.
    (b) The full absorption ${\mathcal P}(\omega)$ obtained by numeric integration of Eq.~(\ref{eq:absorption.w.scattering-general-YaB}). 
    With increasing magnetic field ($\omega_- = 2 H_\perp / \hbar$) the EDSR peaks progressively rise above the ohmic background. Curves are shown for $\omega_- \tau = 12$ (red line), $\omega_- \tau = 24$ (magenta), and $\omega_- \tau = 36$ (blue). Parameters are set to $m/M = 0.1$ and $\mu \tau / \hbar = 50$.
    Inset: the two-peak structure emerges for $\omega_- \tau = 40$ and $\mu \tau / \hbar = 150$. 
    }
    \label{fig:Absorption2}
\end{figure}

In a clean material, $\mu \tau / \hbar \approx k_F l / 2 \gg 1$, 
and visibility of the $\mathcal{P}_\omega^{s}$ peak in 
Eq. (\ref{eq:absorption.w.scattering-general-YaB}) on the background 
of $\mathcal{P}_{\omega}^e$ in Eq. (\ref{eq:Drude}) thus necessarily 
requires $\omega_- \tau \gg 1$.  
That is, the EDSR threshold feature becomes noticeable only when 
it appears in the wings of the Drude Lorentzian in Eq. (\ref{eq:Drude}). 
In Fig. \ref{fig:Absorption2}, we illustrate the evolution of the 
EDSR absorption curve with increasing magnetic field, 
parametrized by $\omega_- \tau$, 
for $\omega_- \tau = 12$, $24$, and $36$. 
Even for the relatively large value $\mu \tau / \hbar \approx k_F l / 2 = 50$,  
at $\omega_- \tau = 12$ the EDSR feature (red line) is barely discernible, 
appearing only as a weak hump on top of the ohmic background. However, with further increasing the field, the onset peak becomes progressively more pronounced, as illustrated by Fig.~\ref{fig:Absorption2}.

Notice that the relaxation times $\tau_s$ and $\tau$, appearing in 
the EDSR peak (\ref{eq:threshold-YaB}) and in the ohmic absorption (\ref{eq:Drude}),  
depend on the scattering mechanisms present in the system and, generally, are different. 
It is thus instructive to corroborate the kinetic-equation treatment above 
by a diagrammatic calculation for Hamiltonian (\ref{eq:H}) in the presence 
of $\delta$-correlated non-magnetic impurities: this will afford us a meaningful 
comparison of the EDSR with the Drude conductivity in the simplest setting, 
where both are shaped by one and the same scattering mechanism. 

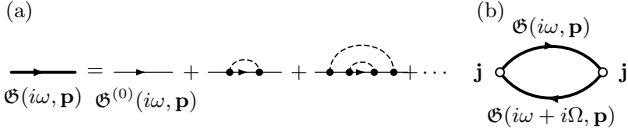
\begin{figure}[ht] 
    \centering
\resizebox{\columnwidth}{!}{
    \begin{tikzpicture}[
        x=1cm,
        y=1cm,
        line cap=round,
        line join=round,
        bare/.style={
            draw,
            line width=0.45pt,
            postaction={decorate},
            decoration={
                markings,
                mark=at position 0.55 with {
                    \arrow{Latex[length=1.5mm,width=1.0mm]}
                }
            }
        },
        dressed/.style={
            draw,
            line width=1.25pt,
            postaction={decorate},
            decoration={
                markings,
                mark=at position 0.55 with {
                    \arrow{Latex[length=1.9mm,width=1.25mm]}
                }
            }
        },
        impurity/.style={
            draw,
            dashed,
            dash pattern=on 2pt off 1.5pt,
            line width=0.55pt
        },
        impurityvertex/.style={
            circle,
            fill=black,
            inner sep=1.1pt
        },
        currentvertex/.style={
            circle,
            draw=black,
            fill=white,
            line width=0.65pt,
            minimum size=3.6pt,
            inner sep=0pt
        },
        every node/.style={
            font=\small
        }
    ]


    \node[anchor=west] at (-0.05,1.70) {(a)};

    \coordinate (d1) at (0.15,0.80);
    \coordinate (d2) at (1.10,0.80);

    \draw[dressed] (d1) -- (d2);

    \node[below=3pt] at ($(d1)!0.5!(d2)$)
        {$\mathfrak{G}(i\omega,\mathbf{p})$};

    \node at (1.40,0.80) {$=$};

    \coordinate (b1) at (1.68,0.80);
    \coordinate (b2) at (2.55,0.80);

    \draw[bare] (b1) -- (b2);

    \node[below=3pt] at ($(b1)!0.55!(b2)$)
        {$\mathfrak{G}^{(0)}(i\omega,\mathbf{p})$};

    \node at (2.83,0.80) {$+$};

    \coordinate (o1) at (3.08,0.80);
    \coordinate (o2) at (4.17,0.80);
    \coordinate (ov1) at (3.40,0.80);
    \coordinate (ov2) at (3.84,0.80);

    \draw[bare] (o1) -- (o2);

    \node[impurityvertex] at (ov1) {};
    \node[impurityvertex] at (ov2) {};

    \draw[impurity]
        (ov1)
        to[out=90,in=90,looseness=1.45]
        (ov2);

    \node at (4.43,0.80) {$+$};

    \coordinate (t1) at (4.67,0.80);
    \coordinate (t2) at (6.05,0.80);

    \coordinate (tv1) at (4.89,0.80);
    \coordinate (tv2) at (5.18,0.80);
    \coordinate (tv3) at (5.55,0.80);
    \coordinate (tv4) at (5.84,0.80);

    \draw[bare] (t1) -- (t2);

    \foreach \p in {tv1,tv2,tv3,tv4}{
        \node[impurityvertex] at (\p) {};
    }

    \draw[impurity]
        (tv1)
        to[out=90,in=90,looseness=1.45]
        (tv4);

    \draw[impurity]
        (tv2)
        to[out=90,in=90,looseness=1.40]
        (tv3);

    \node at (6.33,0.80) {$+\cdots$};


    \begin{scope}[xshift=7.00cm]

        \node[anchor=west] at (-0.05,1.70) {(b)};

        \coordinate (leftvertex)  at (0.42,0.80);
        \coordinate (rightvertex) at (1.95,0.80);

        \draw[dressed]
            (leftvertex)
            to[out=55,in=125,looseness=1.05]
            (rightvertex);

        \draw[dressed]
            (rightvertex)
            to[out=-125,in=-55,looseness=1.05]
            (leftvertex);

        \node[currentvertex] at (leftvertex) {};
        \node[currentvertex] at (rightvertex) {};

        \node[left=4pt] at (leftvertex)
            {$\mathbf{j}$};

        \node[right=4pt] at (rightvertex)
            {$\mathbf{j}$};

        \node[above=11pt] at ($(leftvertex)!0.5!(rightvertex)$)
            {$\mathfrak{G}(i\omega,\mathbf{p})$};

        \node[below=11pt] at ($(leftvertex)!0.5!(rightvertex)$)
            {$\mathfrak{G}(i\omega+i\Omega,\mathbf{p})$};

    \end{scope}

    \end{tikzpicture}
}

    \caption{
    (a) The `rainbow' diagrams, contributing to the 
    disorder-averaged Green's function $\mathfrak{G} (i \omega , {\bf p})$, 
    shown by the bold solid line. Thin solid line denotes the Green's function 
    $\mathfrak{G}^{(0)} (i \omega , {\bf p})$ in an impurity-free crystal: 
    the inverse of the r.h.s. of Eq. (\ref{eq:GMatsubara}) without the last term. 
    Dashed line denotes the disorder correlator 
    $n u_0^2 \delta ({\bf r} - {\bf r}')$, see main text. 
    (b) The polarization `bubble' diagram, corresponding to the Matsubara current-current 
 correlator $\Pi_{\alpha \beta} (i \Omega , {\bf q} = 0)$ in Eq. (\ref{eq:bubble}).
    }
    \label{fig:bubble}
\end{figure}

Such a calculation can be conveniently performed in the Matsubara formalism 
with subsequent analytic continuation to real frequencies. Summation of 
`rainbow' diagrams shown in Fig. \ref{fig:bubble}(a) for uncorrelated 
non-magnetic disorder potential 
$U ({\bf r}) = \sum_i u({\bf r} - {\bf r}_i)$ with zero average 
$\langle U ({\bf r}) \rangle = 0$ and the correlation function 
$\sum_i \langle u({\bf r} - {\bf r}_i) u({\bf r}' - {\bf r}_i) \rangle 
 = n u_0^2 \delta ({\bf r} - {\bf r}')$ produces the textbook self-energy 
 contribution $i \sgn \omega / 2 \tau$ \cite{Abrikosov-Gorkov-1959,Doniach} 
 to the inverse Matsubara Green's function 
\begin{equation}
\label{eq:GMatsubara}
    \mathfrak{G}^{-1} (i \omega , {\bf p}) = 
    i \omega - \xi_{\bf p} - {\bm \Delta}_{\bf p} \cdot {\bm \sigma}
  + {\bf H}_\perp \cdot {\bm \sigma} + \frac{i}{2 \tau} \sgn \omega ,
\end{equation}
with fermion Matsubara frequency $\omega = \pi T (2 l + 1)$, 
and $\tau^{-1} = n |u_0^2| m$, where $n$ is the impurity concentration. 
Calculation of the EDSR absorption amounts to evaluating the polarization operator 
$\Pi_{\alpha \beta} (i \Omega, {\bf q} = 0)$ in Fig. \ref{fig:bubble}(b), with bold lines 
corresponding to the Green's function (\ref{eq:GMatsubara}), 
and the vertices to the current 
${\bf j}^s = - c \sigma_z  \partial \Delta_{\bf p} / \partial {\bf A}$.

For $\delta$-correlated disorder, vertex 
corrections vanish \cite{AGD}, and we only have to evaluate 
\begin{equation}
\label{eq:bubble}
    \Pi_{\alpha \beta}
    (i \Omega) = 
 \tr
 T \sum_{i \omega} \int \frac{d^2 {\bf p}}{(2 \pi \hbar)^2} 
 \left[
 j_\alpha
 \mathfrak{G} (i \omega , {\bf p}) 
 j_\beta
 \mathfrak{G} (i \omega + i \Omega, {\bf p})
 \right] 
\end{equation}
with boson Matsubara frequency $\Omega = 2 \pi T l$. 
As in the kinetic-equation calculation above, Eq. (\ref{eq:bubble}) 
produces two contributions to $\Pi_{\alpha \beta} (i \Omega)$. 
The first contribution, 
\mbox{$\Pi_{\alpha \beta}^e (i \Omega) = \delta_{\alpha \beta} \Pi^e (i \Omega)$},  
arises from both current operators ${\bf j}$ in Eq. (\ref{eq:bubble}) 
being replaced by ${\bf j}^e = \frac{e}{m} {\bf p}$ as in the 
$\delta \mathcal{H}_{\bf p}^e$ of Eq. (\ref{eq:dH}) -- this term yields 
the Drude absorption of Eq. (\ref{eq:Drude}) \cite{Abrikosov-Gorkov-1959,Doniach}. 
The second contribution, 
$\Pi_{\alpha \beta}^s (i \Omega) = \delta_{\alpha \beta} \Pi^s (i \Omega)$, 
arises from both current operators above 
being replaced by ${\bf j}^s = e \sigma_z ({\bm \nabla}_{\bf p} \Delta_{\bf p} )$ 
as in the $\delta \mathcal{H}_{\bf p}^s$ of Eq. (\ref{eq:dH}), 
and it is this very term that accounts for the EDSR. 
With subsequent analytic continuation to real frequencies 
$i \Omega \rightarrow \Omega + i0$, it can be recast as 
\begin{equation} 
\label{eq:bubble-final}
    \Im \Pi^s (\Omega) 
     = 
     - e^2 \frac{m}{M} 
     \int \frac{2 d u}{(2 \pi \hbar)^2} \frac{\omega_-^2}{\omega_{\bf p}^2(u)} 
     \frac{ \Omega / \tau}{[\Omega - \omega_{\bf p}(u)]^2 + \tau^{-2}} ,
\end{equation}
with $\hbar \omega_{\bf p}(u) \equiv 2 \sqrt{u^2 + H_\perp^2}$. 
Together with the relation ${\bf j} = - \frac{1}{c} \Pi {\bf A}$, 
equation (\ref{eq:bubble-final}) allows us to recover the EDSR 
absorption found above using the Boltzmann equation 
-- and thus show that, in the presence of non-magnetic impurities, 
it is the very same scattering time $\tau$ that 
appears in the Drude conductivity and in the EDSR. 




It is important to note that EDSR shall appear not only in altermagnets 
with intrinsic band splitting of Eq. (\ref{eq:H}), but also in 
higher-symmetry N\'eel antiferromagnets, whose electron bands 
are doubly degenerate in the absence of magnetic field.  
In a field ${\mathbf H}_\perp$, applied transversely 
to the staggered magnetization, the symmetry of such antiferromagnets 
remains high enough to protect double degeneracy at certain 
special points in the Brilloin zone \cite{Revaz_2008,Revaz_2009a}. 
At such points, the transverse $g$-factor $g_\perp$ vanishes 
by symmetry, which renders it substantially momentum-dependent 
and turns the Zeeman term into a field-induced spin-orbit coupling 
\begin{equation}
\label{eq:ZeemanSO}
    \mathcal{H}_Z = - 
    \left[
 ({\mathbf H}_\| \cdot {\bm \sigma}) + 
 g_\perp ({\mathbf p}) ({\mathbf H}_\perp \cdot {\bm \sigma})
    \right]
    ] .
\end{equation} 
In a transverse field, the Zeeman splitting thus acquires an `altermagnetic'  
character: already for a crystal of tetragonal symmetry, both $p$- or $d$-wave 
splitting are possible, depending on the location of the band extremum 
in the Brillouin zone \cite{Revaz_2010}. 
This also enables an AC electric field to induce spin resonance transitions 
by virtue of the momentum dependence of $g_\perp ({\mathbf p})$ \cite{Revaz_2009b}. 
While this prediction is yet to be experimentally verified, 
magnetic quantum oscillations observed in two layered antiferromagnetic 
superconductors \cite{RR_Zeeman_SOC_2021} are fully consistent 
with the physics encapsulated in Eq. (\ref{eq:ZeemanSO}).

Note a key difference between the EDSR in such an antiferromagnet 
and that in an altermagnet, described by Hamiltonian (\ref{eq:H}): 
the momentum-dependent splitting 
\mbox{$\hbar \omega_{\bf p} = 2 \sqrt{H_\|^2 + g_\perp^2 ({\mathbf p}) H_\perp^2}$}, 
corresponding to Zeeman term (\ref{eq:ZeemanSO}), now has a flat minimum 
\mbox{$\hbar \omega_- = 2 H_\|$} and a $(\omega - \omega_-)^{- 1/2}$ 
threshold singularity defined by the \textit{longitudinal} 
component $H_\|$ of magnetic field rather than \mbox{$\hbar \omega_- = 2 H_\perp$} 
in an altermagnet. This qualitative difference permits one to use EDSR to 
distinguish an altermagnet from its more familiar antiferrmagnetic counterpart.

To summarize, we predict the effect of electric-dipole spin resonance (EDSR)  
-- \textit{electric} excitation of spin transitions in altermagnetic conductors 
in a transverse magnetic field ${\mathbf H}_\perp$, 
with the lower threshold $\omega_- = 2H_\perp$ of the absorption rate, 
and an inverse square-root singularity $(\omega - \omega_-)^{-1/2}$ 
above this threshold as per Eq. (\ref{eq:main-peak-collisionless}). 
The singularity is cut off by electron scattering, the height of the 
threshold peak scaling as the square root $\sqrt{H_\perp}$ of 
the transverse field as per Eq. (\ref{eq:threshold-YaB}). 
This is to be contrasted with the EDSR in a N\'eel antiferromagnet, 
where the EDSR absorption threshold $\omega_- = 2H_\|$ 
is defined by longitudinal rather than transverse field.

Electric-dipole spin resonance in altermagnets and, more generally, in antiferromagnets, 
thus provides a distinctive spectroscopic signature that both diagnoses altermagnetic order 
and differentiates it from higher-symmetry forms of antiferromagnetism.

We gratefully acknowledge enlightening discussions with M. Kartsovnik, T. Helm and N. Kelly.

\bibliography{Biblio_these}

\end{document}